\documentclass[conference]{IEEEtran}
\IEEEoverridecommandlockouts
\usepackage{cite}
\usepackage{comment}
\usepackage{amsmath,amssymb,amsfonts}
\usepackage{graphicx}
\usepackage{graphicx} 
\usepackage{booktabs}
\usepackage{caption}
\usepackage{multirow}
\usepackage[left=1.62cm,right=1.62cm,top=1.91cm]{geometry}
\usepackage{txfonts}
\usepackage{tikz}
\usepackage{balance}
\usepackage[para]{threeparttable}
\graphicspath{{C:/Users/kiit/Documents/MATLAB/Figures/conf2/}}
\def\BibTeX{{\rm B\kern-.05em{\sc i\kern-.025em b}\kern-.08em
    T\kern-.1667em\lower.7ex\hbox{E}\kern-.125emX}}
    \def\BibTeX{{\rm B\kern-.05em{\sc i\kern-.025em b}\kern-.08em
    T\kern-.1667em\lower.7ex\hbox{E}\kern-.125emX}}
    \makeatletter

\makeatletter
\newcommand{\linebreakand}{%
  \end{@IEEEauthorhalign}
  \hfill\mbox{}\par
  \mbox{}\hfill\begin{@IEEEauthorhalign}
}
\makeatother

\begin{document}

\title{Detection of High Impedance Faults in Microgrids  using Machine Learning}


\author
{\IEEEauthorblockN{Pallav Kumar Bera}
\IEEEauthorblockA{\textit{School of Engr. \& Computer Science} \\
\textit{University of Evansville}\\
Indiana, USA \\
pb141@evansville.edu}
\and
\IEEEauthorblockN{Vajendra Kumar}
\IEEEauthorblockA{\textit{Dept. of Electrical Engineering} \\
\textit{IIT Roorkee , India},\\
kumarvajendra@gmail.com}

\and

\IEEEauthorblockN{Samita Rani Pani}
\IEEEauthorblockA{\textit{School of Electrical Engineering} \\
\textit{VSSUT, Burla, India}\\\textit{KIIT Deemed to be University,  India}\\samita.panifel@kiit.ac.in}

\linebreakand 
\IEEEauthorblockN{Vivek Bargate}
\IEEEauthorblockA{\textit{Dept. of Electrical Engineering} \\
\textit{NIT Raipur, India}\\vicky30786@gmail.com\\
\noindent\makebox[\textwidth][c]{\makebox[11cm]{ \rule{3 cm}{0.5pt}  $ \vardiamondsuit$  \rule{3 cm}{0.5pt}}}
\vspace{-7mm}
}

\thanks{© 2022 IEEE.  Personal use of this material is permitted.  Permission from IEEE must be obtained for all other uses, in any current or future media, including reprinting/republishing this material for advertising or promotional purposes, creating new collective works, for resale or redistribution to servers or lists, or reuse of any copyrighted component of this work in other works}
\thanks{To appear in IEEE Green Energy and Smart Systems Conference,
Long Beach, CA } 
%

}

\maketitle

\begin{abstract}
This article presents differential protection of the distribution line connecting a wind farm in a microgrid. Machine Learning (ML) based models are built using differential features extracted from currents at both ends of the line to assist in relaying decisions. Wavelet coefficients obtained after feature selection from an extensive list of features are used to train the classifiers. Internal faults are distinguished from external faults with CT saturation. The internal faults include the high impedance faults (HIFs) which have very low currents and test the dependability of the conventional relays. The faults are simulated in a 5-bus system in
PSCAD/EMTDC. The results show that ML-based models can effectively distinguish faults and other transients and help maintain security and dependability of the microgrid operation.
\end{abstract}


\section{Introduction}
Microgrids have gathered much attention in the past decade due to the advances in distributed generation (DG) development. 
Microgrid reduces distribution losses, provides generation at the consumption sites, and are more reliable. However, protection of microgrids challenges the conventional relays. The fault current in a microgrid varies depending on the type of DG, operating conditions, and network topology. As a result, traditional overcurrent relays become unreliable \cite{susmita}.
If the faults are high impedance faults (HIFs), the issue becomes even more complex. HIFs that occur in distribution networks are too small in magnitude that they often bypass the conventional relays without getting detected \cite{benner}. The energized conductors in contact with the ground surface are dangerous for animal and plant life and cause losses. HIFs are common and the growth and need for distributed generations will only augment the occurrence of such low current faults \cite{Adamiak}. Hence, a reliable and fast protection system is necessary.

Several strategies for detecting HIFs in distribution networks have been proposed by researchers. 
Mathematical morphology which extracts information from voltage and current was used in \cite{iet}. To detect the low currents associated with HIFs; even, odd, and inter-harmonics in the fault current were used in \cite{ODD, EVEN, INTER}. Genetic algorithm and wavelet transforms were used in \cite{GA} and \cite{wavelet} respectively. Use of Machine Learning (ML) algorithms namely Neural Networks \cite{nn}, Decision Tree \cite{dt}, Support Vector Machines \cite{svm}, k-Nearest Neighbor \cite{knn}, Naive Bayes \cite{bayes} have also been considered. Expert systems have also proven their worth for fault detection and classification in microgrids. An intelligent fault detection
scheme for microgrid systems based on wavelet transform
and deep learning was proposed in \cite{mg2019}. Wavelet transform and decision tree were used for the protection of microgrids with inverter interfaced DGs in \cite{debi}. 
Predictive analytics has also been used to differentiate internal faults and magnetizing inrush in \cite{pallavtx}. Feature-based distance protection was proposed in \cite{pallavwind} for the protection of transmission lines connected to wind farms. This work suggests a comprehensive microgrid protection taking into account various operating modes, including islanded and grid-connected.

There are five sections to the paper. The literature review and overview of this paper are included in Section I. The test system and the creation of the HIF model in PSCAD/EMTDC are described in Section II. The feature extraction and selection are demonstrated in Section III. The findings of this methodology are described in Section IV. Finally, in Section V, conclusions are drawn. 

\section{Modeling}
This section describes the microgrid under study, the modeling of HIF, and power system events simulated in PSCAD/EMTDC considered for the feature extraction and classification.

\subsection{System Studied}
\begin{figure}[ht]
\centering
\includegraphics [width=3.52 in, height= 3.55 in] {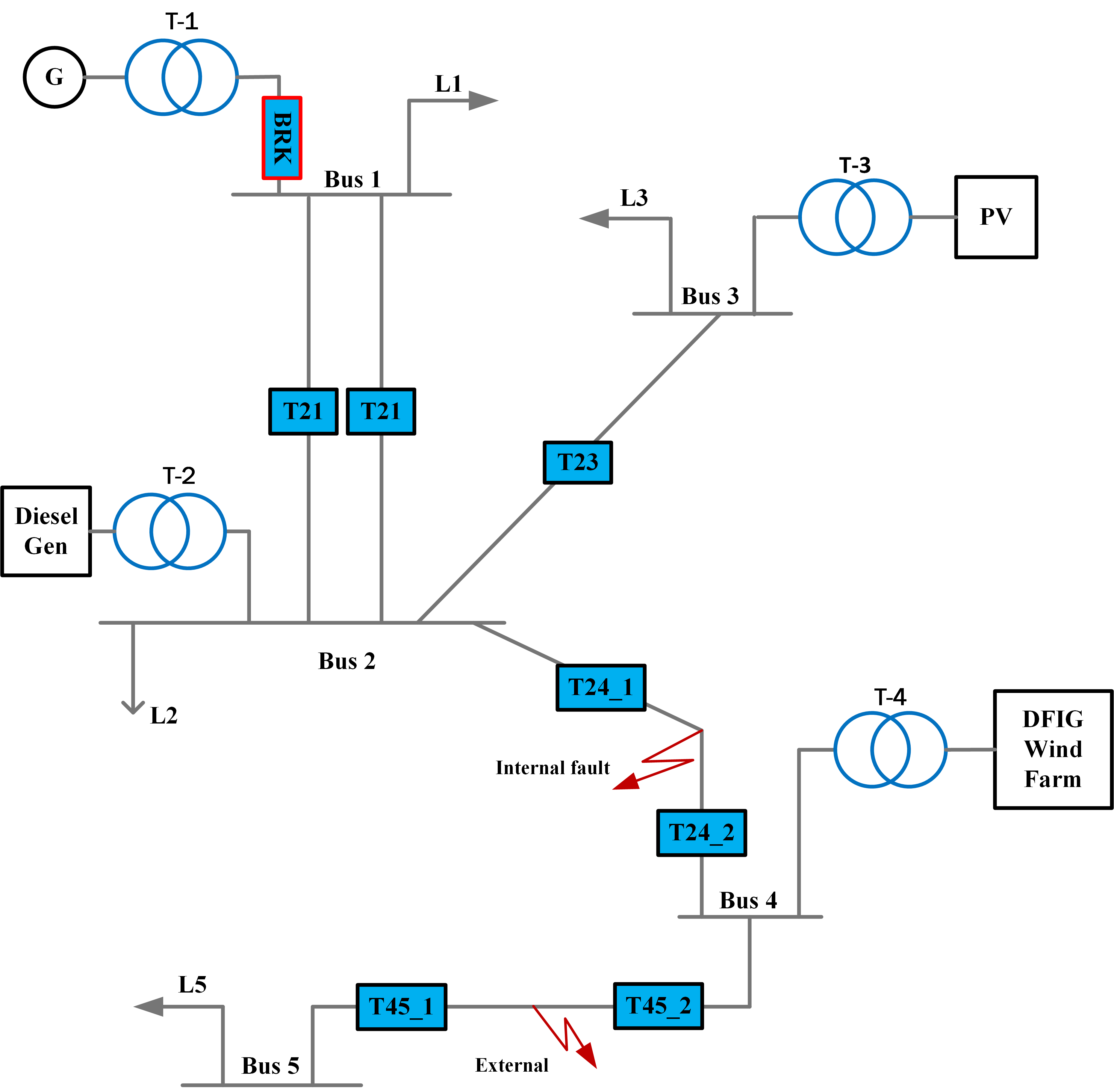} 
\caption{5-bus Test System.}
\label{mg}
\end{figure}

The microgrid system under study is shown in Fig. \ref{mg}. The 5-bus microgrid consists of a Diesel Generator (DG2), PV (DG3), and a Wind Farm (DG4). A DC-DC converter is connected to the PV array.
The PV array's output is determined by irradiance and temperature. The converter adjusts the dc-link current based on the power generated by the maximum power point tracker. The dc voltage is controlled by the voltage source converter. The type-3 wind turbine generator has mechanical and electrical components. The wind turbine extracts the most power possible from the available wind, while the electrical system transforms mechanical energy into electrical energy.  It consists of the induction machine and the AC-DC-AC converter. The governor, IC engine, and synchronous machine are the primary components of a diesel generator.  The details of the distributed generation are given in Table \ref{mg parameters}.

\begin{table}
\centering
\renewcommand{\arraystretch}{1.2}
\setlength{\tabcolsep}{4 pt}
\caption{Parameters of the Microgrid under study}
\label{mg parameters}
\begin{tabular}{l|l} 
\hline\hline
\multicolumn{1}{c|}{\textbf{System}}                                                                                      & \multicolumn{1}{c}{\textbf{ Parameters }}                                                                                                                                                                                                        \\ 
\hline
Main Grid (G)                                                                                                                & Frequency: 60Hz, Rated Voltage: 230 kV                                                                                                                                                                                                           \\ 
\hline
\begin{tabular}[c]{@{}l@{}}Distributed Generations:\\ DG2 (Diesel Generator):\\\\ DG3 (PV)\\ DG4 (Wind Farm)\end{tabular} & \begin{tabular}[c]{@{}l@{}}\\Rated Power:2MW, Rated line voltage:13.8kV \& \\rated RMS line current: 0.833 kA\\Rated Power:0.25MW, at 1000W/m\textsuperscript{2}\& \\nominal temperature of 28\textsuperscript{0} C\\Rated Power: 2.5MW, Wind Speed: 11m/s \end{tabular}  \\ 
\hline
\begin{tabular}[c]{@{}l@{}}Distribution Lines:\\T21\\T23\\T24\\T45\end{tabular}                                           & \begin{tabular}[c]{@{}l@{}}10 km\\20 km\\30 km\\10 km\end{tabular}                                                                                                                                                                               \\ 
\hline
\begin{tabular}[c]{@{}l@{}}Transformers:\\T1\\T2\\T3\\T4\end{tabular}                                                     & \begin{tabular}[c]{@{}l@{}}230kV/ 20kV, 100 MVA~~~~~~~ \\13.86 kV/ 20kV, 3MVA\\460V/ 20kV, 1MVA~~~ \\690V/ 20kV, 5.5 MVA~~~~ \end{tabular}                                                                                                       \\ 
\hline
\begin{tabular}[c]{@{}l@{}}Loads:\\L1\\L2 \\L3\\L5\end{tabular}                                            & \begin{tabular}[c]{@{}l@{}}0.95 MW, 0.475 MVAR\\0.95 MW, 0.475 MVAR\\0.05 MW, 0.025 MVAR\\0.95 MW, 0.475 MVAR\end{tabular}                                                                                                                       \\
\hline\hline
\end{tabular}
\end{table}

\subsection{HIF Modeling}
\begin{figure}[ht]
\centering
\includegraphics [width=1.3 in, height= 2.0 in] {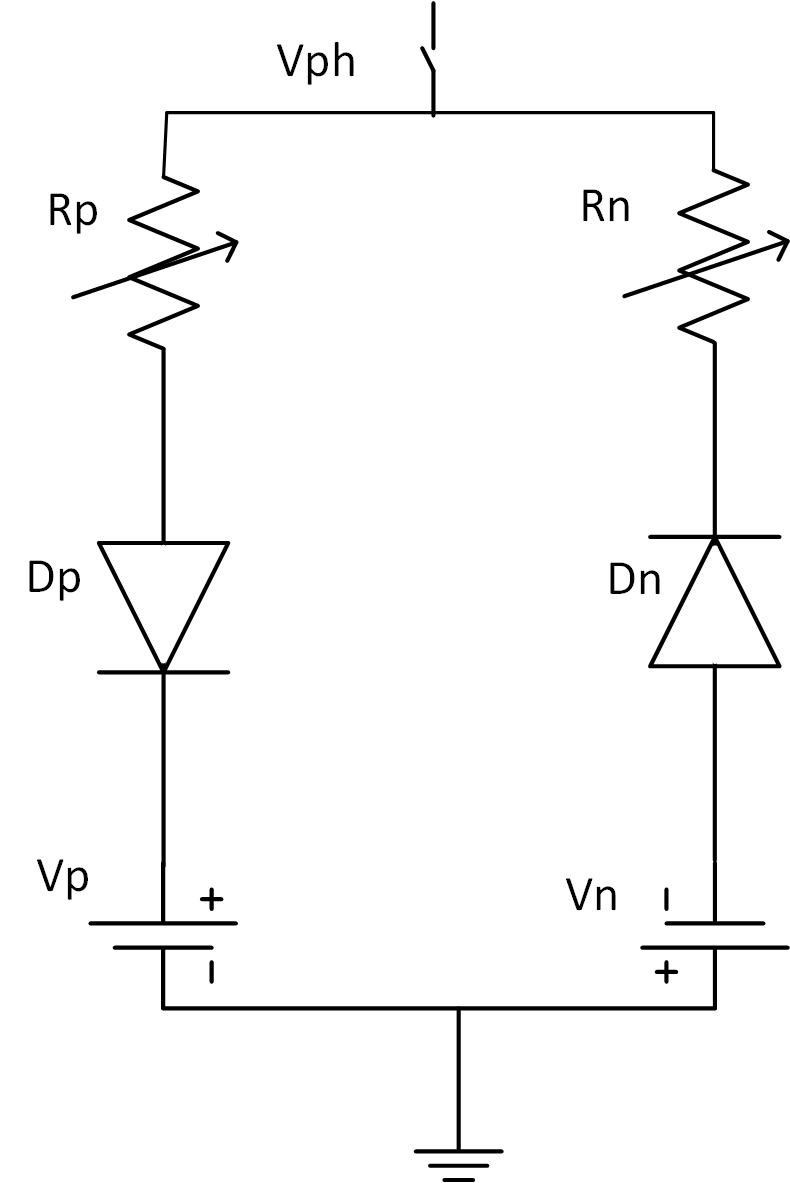} 
\caption{High Impedance Fault model}
\label{hif}

\vspace{3mm}
\end{figure}

There are three primary ways to model HIFs \cite{2019}. Out of these, the one where two- anti-parallel DC sources are connected via two diodes and two variable resistors is used in this paper. 
The HIF model used in this paper is shown in Fig. \ref{hif}. The two variable resistors model the dynamic arc, and the varying DC sources model the asymmetry in the fault currents. In the positive half cycle Vph $>$ Vp,  in the negative half cycle Vph $<$ Vn, and when Vn $<$ Vph $<$ Vp current is zero. 

\begin{table}[ht]
\centering
\renewcommand{\arraystretch}{1}
\setlength{\tabcolsep}{4 pt}
\captionsetup{justification=centering}
\caption{Parameters for External faults}
\label{text}
\begin{tabular}{@{}lcc@{}}
\toprule
\multicolumn{1}{c}{\multirow{2}{*}{\textbf{Parameter \& Values}}} & \multicolumn{2}{l}{\textbf{Operating condition cases}}          \\ \cmidrule(l){2-3} 
\multicolumn{1}{c}{}                                              & \multicolumn{1}{l}{\textbf{Grid Connected}} & \textbf{Islanded} \\ \midrule
Fault Type    (LG, LLG, LL, LLLG, LLL)                            & 5                                           & 5                 \\
Fault Resistance  (0.01, 0.1, 1, 10 $\Omega$s)                    & 4                                           & 4                 \\
Fault Inception Angle  (between 0$^{\circ}$-360$^{\circ}$ )       & 10                                          & 10                \\
Balanced/Unbalanced/Low Voltage                                   & 2                                           & 3                 \\ \midrule
Total cases                                                       & 400                                         & 600               \\ \bottomrule
\end{tabular}
\end{table}

\begin{table}[ht]
\renewcommand{\arraystretch}{1}
\setlength{\tabcolsep}{4 pt}
\captionsetup{justification=centering}
\centering
\caption{Parameters for Internal faults (Non-HIFs)}
\begin{tabular}{@{}lcc@{}}
\toprule
\multicolumn{1}{c}{\multirow{2}{*}{\textbf{Parameter \& Values}}} & \multicolumn{2}{l}{\textbf{Operating condition cases}}          \\ \cmidrule(l){2-3} 
\multicolumn{1}{c}{}                                              & \multicolumn{1}{l}{\textbf{Grid Connected}} & \textbf{Islanded} \\ \midrule
Fault Type  (LG, LLG, LL, LLLG, LLL)                   & 5              & 5        \\
Fault Resistance  (0.01, 0.1, 1, 10, 20 $\Omega$s)           & 5              & 5        \\
Fault Inception Angle (between 0$^{\circ}$-360$^{\circ}$ )                    & 7              & 7        \\
Balanced/Unbalanced/Low voltage & 2              & 3        \\ \midrule
Total   cases                      & 350            & 525      \\ \bottomrule
\end{tabular}
\label{tint}
\end{table}

\begin{table}[ht]
\renewcommand{\arraystretch}{1}
\setlength{\tabcolsep}{4 pt}
\captionsetup{justification=centering}
\centering
\caption{Parameters for HIF faults}
\label{thif}
\begin{tabular}{@{}lcc@{}}
\toprule
\multicolumn{1}{c}{\multirow{2}{*}{\textbf{Parameter \& Values}}} & \multicolumn{2}{l}{\textbf{Operating condition cases}}          \\ \cmidrule(l){2-3} 
\multicolumn{1}{c}{}                                              & \multicolumn{1}{l}{\textbf{Grid Connected}} & \textbf{Islanded} \\ \midrule
Fault Type  (LG)                    & 1            & 1        \\
Phases     (a,b, and c)                   & 3              & 3        \\
Fault Time                & 20             & 20       \\
Balanced/Unbalanced/Low voltage  & 2              & 3        \\ \midrule
Total   cases                      & 120            & 180      \\ \bottomrule
\end{tabular}
\vspace{3mm}
\end{table}

\begin{table}[ht]
\centering

\caption{Events in the Microgrid}
\begin{tabular}{@{}lcc@{}}
\toprule
Event           & Event Type                                                    & No. of Cases                                      \\ \midrule
Internal Faults & \begin{tabular}[c]{@{}l@{}}Type 1 (Non-HIFs) \\ Type 2 (HIFs)\end{tabular} & \begin{tabular}[c]{@{}l@{}}875\\ 300\end{tabular} \\ \midrule
 External Faults & with CT saturation                                            & 1000                                              \\ \bottomrule
\end{tabular}
\label{data}
\end{table}


\begin{figure*}[ht]
\centering
\includegraphics [width=7.2 in, height= 3.5 in] {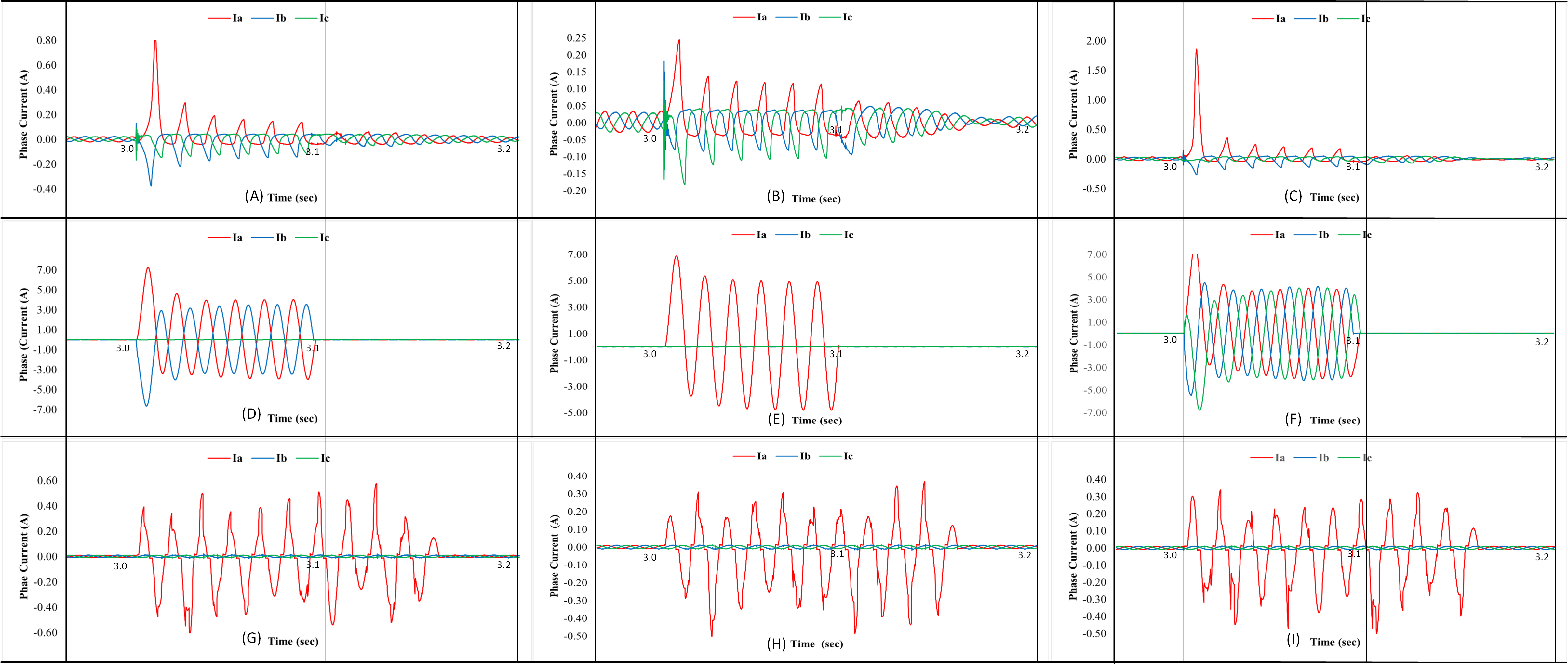}
\caption{External faults with CT saturation (A,B,C), internal faults (D,E,F), and HIF faults (G,H,I)}
\label{faults}
\vspace{3mm}
\end{figure*}

\begin{table*}
\renewcommand{\arraystretch}{1.2}
\setlength{\tabcolsep}{4 pt}
\centering
\caption{Hyperparameter Selection}
\label{hs}
\begin{tabular}{l|l|l} 
\hline
\hline
\textit{\textbf{Classifier}} & \textit{\textbf{Hyperparameter grid}}                                                                                                                                                                                                              & \textit{\textbf{Best Hyperparameters}}                                                                                                                                                       \\  \hline
Decision Tree                & \textbf{criterion:} entropy,gini                                                                                                                                                                                                                   & \textbf{criterion:} entropy                                                                                                                                                                  \\
Random Forest                & \begin{tabular}[c]{@{}l@{}}\textbf{max\_depth:} 4-20, \textbf{max\_features:} 0.1-1, \textbf{}\\\textbf{min\_samples\_leaf:} 5-60,\textbf{ estimators:} 500-4000\end{tabular}                                                                      & \begin{tabular}[c]{@{}l@{}}\textbf{max\_depth:}5, \textbf{max\_features:} 0.1, \textbf{min\_samples\_leaf:} 5,\\\textbf{estimators:} 500\end{tabular}                                        \\
Gradient Boost               & \begin{tabular}[c]{@{}l@{}}\textbf{learning\_rate:}0.01-0.1, \textbf{max\_depth:} 4-20, \textbf{}\\\textbf{estimators:} 500-4000,\textbf{ subsample:} 0.5-1\end{tabular}                                                                           & \begin{tabular}[c]{@{}l@{}}\textbf{learning\_rate:}0.01, \textbf{max\_depth:} 5, \textbf{estimators:} 500,\\\textbf{subsample:} 0.5\end{tabular}                                             \\
Multi Layer Perceptron       & \begin{tabular}[c]{@{}l@{}}\textbf{activation:} logistic, identity, tanh, relu, \textbf{alpha:} 0.0001-0.01, \textbf{}\\\textbf{hidden\_layer:} (30-100, 8-30), \textbf{learning\_rate:} adaptive,\\\textbf{solver:} lbfgs, adam, sgd\end{tabular} & \begin{tabular}[c]{@{}l@{}}\textbf{activation:} logistic, \textbf{alpha:} 0.0001, \textbf{hidden\_layer:} (88, 23),\\\textbf{learning\_rate:} adaptive, \textbf{solver:} lbfgs\end{tabular}  \\
Naive Bayes                  & -                                                                                                                                                                                                                                                  & -                                                                                                                                                                                            \\
K-Nearest Neighbor           & \begin{tabular}[c]{@{}l@{}}\textbf{leaf\_size:} 2-15, \textbf{neighbors:} 3-10, \textbf{}\\\textbf{distance:} manhattan, euclidean\end{tabular}                                                                                                    & \textbf{leaf\_size:}3, \textbf{neighbors:} 4, \textbf{distance:} manhattan                                                                                                                   \\
Support Vector Machine       & \textbf{C:}0.1-1000, \textbf{gamma:} 0.0001-1, \textbf{kernel:} rbf, poly                                                                                                                                                                          & \textbf{C:}1000, \textbf{gamma:} 1, \textbf{kernel:} rbf                                                                                                                                     \\
\hline
\hline
\end{tabular}
\vspace{3mm}
\end{table*}
\subsection{Power System Events}
The distribution line T24 between bus-2 and bus-4 is considered for simulating the internal faults which include the HIFs. Type 1 internal faults include  LG, LLG, LL, LLLG, and LLL faults, whereas Type 2 internal faults, also referred to as the HIFs, are the downed
conductor faults for each phase. Tables \ref{text}, \ref{tint}, and \ref{thif} show the parameters (fault type, fault resistance, fault inception angle, etc.), their values, and the number of cases simulated for the fault data in grid-connected and islanded modes. For internal and external faults, balanced and unbalanced loads with LG, LLG, LL, LLL, and LLLG fault types are evaluated, whereas only LG faults in the three phases for HIFs are considered. For HIFs the fault impedance is varied between 50 $\Omega$ to 300 $\Omega$ randomly in the interval of 0.2ms during fault. Lower voltages up to 0.9 per unit are also taken into account while simulating the faults \cite{hosiyar}. The external faults are simulated in line T45 while the CTs at the two ends of line T24 were burdened differently. Fig. \ref{faults} shows the 3-phase differential currents for external faults with CT saturation, internal faults, and high impedance faults. (A), (B), and (C) show external faults with CT saturation; (D), (E), and (F) show internal Type 1 faults; and (G), (H), and (I) show internal Type 2 faults in grid-connected, islanded, low voltage condition, balanced and unbalanced loading conditions.
Table \ref{data} shows the number of internal faults and external faults simulated in the microgrid. Using the parameters mentioned above, 1175 internal faults and 1000 external faults are generated.

\section{Feature extraction and selection}
Feature selection helps identify key characteristics in the differential currents. The selection of informative, discriminating, and independent features is a critical component in pattern recognition and classification tasks. It also helps to reduce the input matrix size and thus increases the credibility of the model. In \cite{pallavswing, pallavsystem, pallavaccess} several time, time-frequency, and frequency domain features were extracted and then selected to detect and classify power system events like faults and transients. Here, the importance of individual features is calculated in terms of information gain. The features with higher information gain are more relevant in classifying the faults and non-fault events.

The features which can be spectral (FFT coeff. etc.), time-frequency (wavelet coeff.), information theory (entropy, approximate entropy, etc.), or statistical (autocorrelation, minimum, maximum, mean, median, etc.)  are calculated in Python. The list of features is detailed in \cite{tsfresh}.
2289 features are examined and then the topmost features from the 3-phase differential currents obtained via the CTs at both ends are used to train the classifier models. Thus, Continuous Wavelet coefficients are obtained in the feature selection process.

\subsection{Continuous Wavelet Transform}
The Mexican Hat wavelet, $\psi$(t, p, q) is obtained by scaling and shifting the mother wavelet  $\psi$(t)  \begin{equation}\label{mhw}
\psi(t,p,q) =\frac{2}{\sqrt{3p} \pi^{\frac{1}{4}}} (1 - \frac{(t-q)^2}{p^2}) exp(-\frac{(t-q)^2}{2p^2})\end{equation} 
where p and q are the scaling and shifting parameters.

The Continuous Wavelet Transform (CWT) of a signal y(t) is then obtained by using
\begin{equation}\label{cwt}
Y(y(t),p,q,\psi(t)) = \int_{- \infty}^{+\infty}y(t) \psi^*(t,p,q) dt\end{equation}
where $\psi^*(t,p,q)$ is the complex conjugate of  $\psi(t,p,q)$
\vspace{3mm}

\begin{table}[ht]
\renewcommand{\arraystretch}{1.2}
\setlength{\tabcolsep}{4 pt}
\caption{Model Performance}
\label{result}
\begin{tabular}{@{}l|c|c|c@{}}
\hline
\hline
\textbf{Classifier}             & \textbf{Balanced Accuracy} & \textbf{Dependability} & \textbf{Security} \\ \hline
Decision Tree          & 99.68\%           & 99.72\%       & 99.64\%  \\
Random Forest          & 99.23\%           & 99.17\%       & 99.29\%  \\
Gradient Boost         & 99.68\%           & 99.72\%       & 99.64\%  \\
Multi Layer Perceptron & 99.85\%           & 99.7\%        & 100\%    \\
Naive Bayes            & 56.3\%            & 12.6\%        & 100\%    \\
K-Nearest Neighbor     & 99.86\%           & 99.72\%       & 100\%    \\
Support Vector Machine & 100\%             & 100\%         & 100\%    \\ \hline
\hline
\end{tabular}
\end{table}

\section{Results}
The balanced accuracy (eq.\ref{accbar}) which is a better representation of accuracy is obtained for all the classifiers considered in the study. Eq. \ref{d} is used to calculate the dependability for internal faults, which includes HIF faults, and eq. \ref{s} is used to calculate the security for external faults with CT saturation. 
 \begin{equation}
\label{accbar}\bar{\eta}=\frac{1}{2}(\frac{TP}{TP+FN}+ \frac{TN}{TN+ FP})\end{equation} 
 \begin{equation}
\label{d}{Dependability}=(\frac{TP}{TP+FN})\end{equation}
 \begin{equation}
\label{s}{Security}=(\frac{TN}{TN+FP})\end{equation}
Here, TP is true positive, TN is true negative, FP is false positive, and FN is false negative. The classifier models used to differentiate the internal faults and external faults are: Decision Tree, Random Forest, Gradient Boost, K-Nearest Neighbor, Naive Bayes, Support Vector Machine, and Multi Layer Perceptron.
It is critical to find the optimal hyperparameters (parameter that affects how the learning process is carried out) for these algorithms. It fine-tunes the parameters and minimizes the loss function. The hyperparameters are optimized using grid search over possible values of the individual parameter. Table \ref{hs} shows the hyperparameter grid and the best choice for testing the different classifiers considered in this study.
Cross-validation is used to avoid overfitting and to credibly measure the merit of the prediction. The classification results are given in Table \ref{result}. The Wavelet transform coefficients obtained from the external and internal faults simulated in Section II are used for training and testing the seven classifiers. On this fault data, all classifiers except Naive Bayes fared well, with Support Vector Machine outperforming the others. 
\vspace{3mm}

\section{Conclusion}
The paper presents differential protection of lines connecting a wind farm in a microgrid. The differential features obtained from the currents at both ends of the line under consideration are used to build the ML-based models. 2175 external and internal faults conditions are simulated by varying different system parameters. The findings suggest that models based on ML can successfully differentiate internal and external faults. The performance of the ML algorithms is analyzed in the presence of high impedance faults. ML-based fault detection is able to identify internal and external faults with CT saturation even in the face of these low-magnitude current faults.

\balance

\bibliography{ref} 
\small{
\bibliographystyle{ieeetr}
}
\noindent\makebox[\linewidth]{       ------------------*------------------}
\end{document}